# Phonons and Oxygen Diffusion in $Bi_2O_3$ and $(Bi_{0.7}Y_{0.3})_2O_3$

Prabhatasree Goel[1], M. K. Gupta[1], R. Mittal[1,2,*], S. J. Skinner[3], S. Mukhopadhyay[4,$]
S. Rols[5] and S. L. Chaplot[1,2]
[1]*Solid State Physics Division, Bhabha Atomic Research Centre, Trombay, Mumbai 400085, India*
[2]*Homi Bhabha National Institute, Anushaktinagar, Mumbai 400094, India*
[3]*Department of Materials, Imperial College London, London, SW7 2AZ, UK*
[4]*ISIS Neutron and Muon Source, Rutherford Appleton Laboratory, Didcot, Oxon OX11 0QX, UK*
[5]*Institut Laue-Langevin, BP 156, 38042 Grenoble Cedex 9, France*
Corresponding Authors Email: rmittal@barc.gov.in[*], sanghamitra.mukhopadhyay@stfc.ac.uk[$]

We report investigation of phonons and oxygen diffusion in $Bi_2O_3$ and $(Bi_{0.7}Y_{0.3})_2O_3$. The phonon spectra have been measured in $Bi_2O_3$ at high temperatures up to 1083 K using inelastic neutron scattering. Ab-initio calculations have been used to compute the individual contributions of the constituent atoms in $Bi_2O_3$ and $(Bi_{0.7}Y_{0.3})_2O_3$ to the total phonon density of states. Our computed results indicate that as temperature is increased, there is a complete loss of sharp peak structure in the vibrational density of states. Ab-initio molecular dynamics simulations show that even at 1000 K in δ-phase $Bi_2O_3$, Bi-Bi correlations remain ordered in the crystalline lattice while the correlations between O-O show liquid like disordered behavior. In the case of $(Bi_{0.7}Y_{0.3})_2O_3$, the O-O correlations broadened at around 500 K indicating that oxygen conductivity is possible at such low temperatures in $(Bi_{0.7}Y_{0.3})_2O_3$ although the conductivity is much less than that observed in the undoped high temperature δ-phase of $Bi_2O_3$. This result is consistent with the calculated diffusion coefficients of oxygen and observation by QENS experiments. Our ab-initio molecular dynamics calculations predict that macroscopic diffusion is attainable in $(Bi_{0.7}Y_{0.3})_2O_3$ at much lower temperatures, which is more suited for technological applications. Our studies elucidate the easy directions of diffusion in δ-$Bi_2O_3$ and $(Bi_{0.7}Y_{0.3})_2O_3$.

## I. INTRODUCTION

Solids exhibiting high levels of oxygen ion conduction have always been of considerable interest to researchers all over the world[1-7]. These solids find potential technological applications in fuel cells, gas sensors, ceramic oxygen generators and are the backbone for light photo-catalysts and redox catalysts for several chemical processes [3,4,7-11]. High ionic conductivity in these compounds is attributed to the migration of oxygen-ion vacancies at high temperature. Yttria-stabilized zirconia (YSZ), a widely used oxygen-ion conductor, is a typical example. In this category, bismuth sesquioxide ($Bi_2O_3$) is a frontrunner, which has a conductivity two orders of magnitude greater[12,13] than YSZ at temperatures above 1000 K. $Bi_2O_3$ occurs in the monoclinic α-phase at ambient conditions[3,4,14], whilst the high temperature δ-phase above 1003 K exhibits defective fluorite structure including a large amount of vacant sites in the oxide ion sublattice[14,15]. Also, $Bi_2O_3$ is polymorphic, depending upon temperature and thermal history; several phases like β, γ, δ, ε and ω phases are formed. The defect fluorite type structure[16] (δ-phase) of $Bi_2O_3$ has six anions ($O^{2-}$) and they are randomly distributed over the eight tetrahedral holes of the fcc lattice formed by $Bi^{3+}$ ions. The δ-phase is known[2,13] to have high solid-state oxide ionic conductivity (1–2 S cm$^{-1}$) over its narrow stability range of 1003−1090 K. The fast high oxygen ion conduction behavior of δ-$Bi_2O_3$ in general is linked[17] to the intrinsic oxygen ion vacancies and asymmetric arrangements of anions due to the lone pair ($6s^2$) of $Bi^{3+}$ ions.

There have been several theoretical and experimental studies to understand the origin of diffusion in this oxide. Several structural models have been formulated to understand the high temperature defective cubic phase of $Bi_2O_3$. Gattow and Schroder suggested[3,4,16,18,19] a random distribution of oxygen ions over the tetrahedral positions, so there is an equal probability of oxygen occupying any of the Wyckoff sites. Battle et al[20] proposed that 32f interstitial sites are occupied, Sillen[3,4,18,21] et al surmised that three-quarters of regular tetrahedral sites are occupied while one-quarters are vacant, leaving a lattice with 25% oxygen vacant contents. Willis[3,4,18] et al suggested that the oxygen ions are displaced along <111> directions from their regular sites towards the octahedral vacancies. Several other authors suggest a preference along <111> directions of the oxygen atoms, while there are inferences of a preference for <110> as it is energetically more favourable. Studies using density functional theory (DFT) have shown that <100> oxygen vacancy alignment has the lowest energy while along <110> configuration it is marginally higher. Mohn et al reported[3] that the irregular local structure of δ –$Bi_2O_3$ is closely connected to increased electron density around Bi, thus facilitating the presence of a stereo chemically active lone-pair. This observation of higher cation polarizability associated with the lone pair of electrons in Bi, is understood to be a key factor in achieving sustained oxygen diffusion. Neutron scattering studies[15] reveal the presence of micro domains or short range order where in there is local ordering. These various studies have tried to understand the role of local structure disorder and its electronic properties in diffusion. With aging, oxygen ions slowly undergo some ordering processes of the unoccupied sites leading to reduction in the conductivity of $Bi_2O_3$. There have been diffuse elastic neutron scattering studies[22] on doped $Bi_2O_3$ to understand the effect of different cation doping on structure of the cubic phase.

Ionic conductivity of $Bi_2O_3$ drops at room temperature. The problem of stabilizing the high conducting phase at room temperature can be eliminated by suitable cation doping, such as rare-earth ions, $Y^{3+}$, $Sc^{3+}$, etc. The high ionic conducting δ-phase can be stabilized[6,23] at ambient temperature in a range of compositions, viz., $(Bi_{1-x}Y_x)_2O_3$ (x=0.1 - 0.5). Doping with suitable cations stabilizes the superionic conducting phase of $Bi_2O_3$ at lower temperature, but the flip side is that it also reduces the ionic conductivity as the Arrhenius energy for oxygen migration is raised. The ionic conductivities of these non-stoichiometric compounds are found to be higher than that of yttria-stabilized zirconia at similar temperature. The studies on diffusion[19,24] and transport properties of yttria-doped bismuth oxide shows that it can be a potential material for low or intermediate temperature applications based on oxygen ion diffusion.

Progress is being made to achieve solid electrolytes with ionic conductivity of the order of 0.0001 to 0.1 $Ohm^{-1}$ $cm^{-1}$ and activation energy of about 0.1 eV. Accurate measurement and calculation of phonon frequencies are needed to evaluate the jump frequency and hence the diffusion coefficient. Quasielastic neutron scattering (QENS) is a very good technique to study[25-27] oxygen ion transport in such oxides. The diffusion of oxygen atoms can be investigated using this technique to understand the microscopic nature of diffusion, such as, the jump length, jump path and the residence time, etc. Recently we have employed ab-initio DFT and molecular dynamics (MD) simulations in the analysis of inelastic neutron scattering (INS) measurements[28-31], Ab-initio MD (AIMD) also been used to analyse QENS experiments to understand diffusions microscopically[32,33].

Here we report detailed vibrational study on the different phases of the $Bi_2O_3$ and $(Bi_{0.7}Y_{0.3})_2O_3$. Further, we have performed extensive ab-initio molecular dynamics simulations to study the phonons and diffusion of oxygen in δ-$Bi_2O_3$ and in 30% $Y_2O_3$ doped in $Bi_2O_3$, i.e., $(Bi_{0.7}Y_{0.3})_2O_3$. We also report neutron inelastic measurements of the phonon density of states in $Bi_2O_3$ at different temperatures, from 300 K to 1048 K. Subsequently QENS measurements are done in the temperature range 353 K - 1083 K. Possible preferred direction of oxygen ion movement has always been of intrigue and several studies have been devoted to understand the same. Our aim is to understand the difference in oxygen ion diffusion in the δ-$Bi_2O_3$ and $(Bi_{0.7}Y_{0.3})_2O_3$ by employing an integrated experimental, and ab-initio DFT and MD based simulations

## II. EXPERIMENTAL

Samples of $Bi_2O_3$ were obtained from Sigma Aldrich, UK (99.999%) and used without further processing. The inelastic neutron scattering experiment on $Bi_2O_3$ was carried out using the IN4C spectrometer at the Institut Laue Langevin (ILL), France. The spectrometer was based on the time-of-flight (TOF) technique and was equipped with a large detector bank covering a wide range of about 10° to 110° of scattering angle. The INS measurements were performed at several temperatures from 300 K to 1048 K. About 2 cc of polycrystalline sample of $Bi_2O_3$ has been used for the measurements. For these measurements we have used an incident neutron wavelength of 2.4 Å (14.2 meV) in neutron energy gain setup. In the incoherent one-phonon approximation, the measured scattering function $S(Q,E)$, where E and Q were the energy transfer and momentum transfer vector, respectively, was related[34-36] to the phonon density of states $g^{(n)}(E)$ as follows:

$$g^{(n)}(E) = A \left\langle \frac{e^{2W(Q)}}{Q^2} \frac{E}{n(E,T) + \frac{1}{2} \pm \frac{1}{2}} S(Q,E) \right\rangle \quad (1)$$

$$g^n(E) = B \sum_k \left\{ \frac{4\pi b_k^2}{m_k} \right\} g_k(E) \quad (2)$$

where the + or – signs correspond to energy loss or gain of the neutrons, respectively, $n(E,T) = [\exp(E/k_B T) - 1]^{-1}$, T was temperature and $k_B$ Boltzmann's constant. *A* and *B* were normalization constants and $b_k$, $m_k$, and $g_k(E)$ were, respectively, the neutron scattering length, mass, and partial density of states of the $k^{th}$ atom in the unit cell. The quantity between < > represented suitable average over all *Q* values at a given energy. $2W(Q)$ was the Debye-Waller factor averaged over all the atoms. The weighting factors $\frac{4\pi b_k^2}{m_k}$ for various atoms in the units of barns/amu were: 0.0438 and 0.2645 for Bi and O respectively.

The QENS experiments were performed on the OSIRIS spectrometer[37] of the ISIS Neutron and Muon source, UK. This instrument was a high-resolution indirect-geometry time-of-flight backscattering spectrometer with final energy of $E_f$ = 1.845 meV at the analyser setting PG002. In this setting pyrolithic graphite analysers used reflection at its 002 plane to determine the final neutron energy given above and energy resolution 27 µeV. A quartz sample cell was filled with $Bi_2O_3$ powder and attached with a gas handling rig supplying pure oxygen. Over the high temperature measurements, the $O_2$ pressure was maintained at 240 mbar and data collected for 5 hours at each temperature. Prior to data collection a series of measurements were taken on the empty quartz ampoule. Initial data from the empty quartz were obtained at 353 K, before gathering data every 20 K from 1023 K – 1083 K. In this article the analysis of the result obtained at 1083 K has been presented.

The data analysis was undertaken using the QENS data analysis interface as implemented in the Mantid software[38]. One delta function and one Lorentzian together were convoluted with instrument resolution determined from the Vanadium standard cell at room temperature. The background data of the quartz cell has been subtracted from all data analysis mentioned below.

## III. COMPUTATIONAL DETAILS

The lattice and molecular dynamics simulations were performed in the ordered[39] (monoclinic α-phase) and disordered δ-phase[16] of $Bi_2O_3$ (cubic phase, space group Fm-3m) and $(Bi_{0.7}Y_{0.3})_2O_3$ (cubic phase, space group Fm-3m) using ab-initio DFT as implemented in the VASP simulation package[40,41].

A supercell of (2 × 2 × 2) dimension, which consist of 160 atoms, has been used in the computations of the ordered[39] monoclinic α-phase. In the lattice dynamics calculations, the required force constants were computed within the Hellman-Feynman framework, on various atoms in different configurations of a supercell with (±x, ±y, ±z) atomic displacement patterns. The generalized gradient approximation (GGA) exchange correlation following the parameterization by Perdew, Becke and Ernzerhof[42,43] has been used for the computation of total energy and forces using the projected augmented wave (PAW) formalism of the Kohn-Sham density functional theory. An energy cut-off of 900 eV was used for plane wave expansion. The Monkhorst Pack method[44] was used for k point generation with a 4×4×4 k-point mesh. The convergence breakdown criteria for the total energy and ionic force loops were set to $10^{-8}$ eV and $10^{-4}$ eV $\text{Å}^{-1}$, respectively. We have used PHONON software[45] to obtain the phonon frequencies in the entire Brillouin zone, as a subsequent step to density functional theory total energy calculations.

The ab-initio molecular dynamics (AIMD) simulations were performed in the NVE ensemble for 60 pico-second, with a time step of 2 femtosecond. A supercell of (2 × 2 × 2) dimension has been used. The AIMD simulation supercell of monoclinic α-phase consist of 160 atoms while disordered δ-phase[16] of $Bi_2O_3$ (cubic phase, space group Fm-3m) and $(Bi_{0.7}Y_{0.3})_2O_3$ have 80 atoms. We have taken a single k-point in the Brillouin zone and an energy cut-off of 900 eV was used for plane wave expansion. The energy convergence criteria of $10^{-5}$ eV has been chosen for self-consistence electron energy convergence. Initially, the structure was equilibrated for 10 picoseconds to attain the required temperature in NVT simulations. The temperature in the NVT simulations was attained through a Nose thermostat[46]. Then the production run was performed for up to 60 picoseconds within NVE ensemble. Simulations were performed for a series of temperatures from 300 to 1100 K. At each temperature, a well-equilibrated configuration was obtained during the whole simulation. At 1100 K, the simulations were extended up to 200 picoseconds.

The dynamical structure factor S(Q,E) were calculated from the trajectory obtained from AIMD simulations using nMoldyn software following the procedure reported previously[33,47].

## IV. RESULTS AND DISCUSSIONS

### A. Temperature Dependence of Phonon Density of States

We have performed INS measurements of $Bi_2O_3$ at several temperatures (**Fig. 1(a)**) from 300 K to 1048 K beyond the superionic transition at about 1003 K. The room temperature measurements show well defined peak structure in the phonon density of states which disappears at above 973 K below the superionic transition temperature. We have compared the measured phonon density of states at 300 K with the lattice dynamics calculated phonon spectrum in the ambient α-phase (**Fig. 1(b)**). The neutron-weighted phonon density of states shows peaks at about 7, 15, 25, 40, 50 and 61 meV. The calculated total neutron-weighted phonon density of states is in good agreement with measurements (**Fig 1(b)**). The calculated partial contributions from Bi and O atoms to the total phonon density of states are also shown. It can be seen that due to its heavier mass Bi contributes mainly at low energy up to 20 meV, while the contribution from O is in the entire spectrum. All the peaks in the experimental phonon spectra agree well with our calculations. As temperature increases to 773 K, peaks at about 25, 40, 43, 51, and 60 meV all are reduced in intensity. At 973 K, peak at 25 meV almost disappears leaving a broad peak centered around 48 meV. Beyond 973 K, the spectrum lost all the sharp peak structure in the phonon density of states (see Fig 1(a)).

To understand the atomistic contributions in the vibrational densities of states (VDOS), partial VDOS of Bi and O are given in Fig. 1(b). To include anharmonic contributions in the VDOS, which may be relevant at high temperature we performed AIMD simulation to evaluate the phonon spectrum. This has been calculated[48,49] using the Fourier transform of velocity auto-correlation function- obtained from AIMD simulations. As given below our calculations are able to reproduce the experimental spectrum fairly well.

In **Fig 2** we have compared the calculated partial VDOS of Bi, O in the α- and the δ-phase of $Bi_2O_3$ as well as in $(Bi_{0.7}Y_{0.3})_2O_3$ at different temperatures using AIMD simulations. In the ambient α-phase, Bi has contributions up to 25 meV at 300 K; with increase in temperature to 800 K, the Bi contribution remains unchanged, however, the oxygen contribution in the entire Brillouin zone shifts towards lower energies. This shift can be seen from the total VDOS (**Fig. 2**). In the highly conducting δ-phase, the partial contribution of Bi does not show any significant change and here again oxygen contributes in the entire Brillouin zone. The partial and total density of states at 1000 K and 1100 K for the δ phase do not show much change. In case of $(Bi_{0.7}Y_{0.3})_2O_3$, the computed spectra have been plotted for 300 K, 1000 K and 1100 K. Bi contribution is found to be same as that in $Bi_2O_3$, i.e up to 25 meV, while Y contributes up to 40 meV. This is in accordance to their masses; Y's mass is 88.9 amu while Bi's is 208.9 amu. Oxygen (16 amu) contributes in the entire range. With increase in temperature, the spectra move down (**Fig. 2**) in the energy as expected. The data at 1000 K and 1100 K does not show much change.

The comparison between the molecular dynamics results with the measured phonon density of states is given in **Figure 3(a) and 3(b)**. The ambient phase experimental data have been compared with the computed results and it is found that although the calculations of the α-phase at 300 K do not show distinct peak structure, but peaks at 5, 25, 48, 54 and 60 meV compare well with the experimental data. Similarly, the data at 1100 K of the δ-phase is in qualitative agreement with the experimental data at 1048 K.

We found a good agreement between the calculation and measurements which validate the employment of the AIMD method for further microscopic analysis of the transport properties of the pure and doped oxide of Bi. The phonon density of states calculations performed using both lattice dynamics (**Fig. 1(b)**) and molecular dynamics simulations (**Fig. 3**) show a satisfactory agreement indicating a good degree of convergence.

**B**. **Pair Correlation Function**

In order to understand the microscopic picture of the lattice at ambient and at higher temperature in ordered $Bi_2O_3$, disordered $δ-Bi_2O_3$ and $(Bi_{0.7}Y_{0.3})_2O_3$, we have plotted pair correlations of different pair of atoms in **Fig 4**. In the ambient temperature of α- phase at 300 K, the minimum Bi-Bi distance is 3.5 Å, while the nearest neighbor Bi-O distance is approximately 2.5 Å. and the closest O-O distance is 3.2 Å. With increase in temperature, at 800 K there are no discernable changes in the pair correlations in $α-Bi_2O_3$. On further increase in temperature to 1000 K, in the cubic $δ-Bi_2O_3$ phase, however, there is definite correlation between Bi-Bi. The correlation between Bi-O is less significant after the first neighbor, while for O-O it is broadened considerably at both 1000 K and 1100 K. In these temperatures, the nearest neighbor distance for Bi-Bi is 4 Å. In the case of Bi-O it is 2.4 Å while for O-O it is about 3 Å. This shows the increased disturbances occurring in the oxygen sub-lattice at this temperature range.

In the case of (Bi$_{0.7}$Y$_{0.3}$)$_2$O$_3$, however, there is definite correlation between Bi-O and Bi-Y up until the second neighbor at different temperatures. Bi-Bi correlations start at 4 Å, which is the same for Y-Y. The first nearest distance for Bi-O and Y-O is the same at 2.2 Å. Bi-Y correlations start at 4 Å, while for O-O the first neighbor distance is about 3 Å at 300 K. Moreover, in the case of O-O correlation, it is very poor even at 300 K. There are no changes in the Bi-Bi, Bi-Y, Bi-O correlations as temperature is increased from 300 K to 1100 K. Our results predict that the first peak in the correlation functions, corresponding to Bi-O nearest neighbor distance, in the disordered phase is around 2.2 Å which is in very good agreement with Mohn et al's ab-initio MD results[3]. Both in cubic disordered Bi$_2$O$_3$ and doped Bi$_2$O$_3$ the correlations are broad after this first peak while in the case of the α-phase there is definite peak structure beyond the first neighbor distance. The Bi-Bi first neighbor distance in the α phase is 3.5 Å while that in the case of both doped and disordered δ-phase is about 4 Å.

## C. Mean Squared Displacement and Diffusion

The mean square displacement (MSD) of various atoms at time $\tau$ is calculated using the following relation[50,51]

$$u^2(\tau) = \frac{1}{N_{ion}(Nstep-N\tau)} \sum_{i=1}^{N_{ion}} \sum_{j=1}^{N_{step}-N\tau} |r_i(t_j+\tau) - r_i(t_j)|^2 \qquad (3)$$

Here $r_i(t_j)$ is the position of i$^{th}$ atom at j$^{th}$ time step. N$_{step}$ is the total number of simulation steps and N$_{ion}$ is the total number of atoms of a particular type in the simulation cell. N$\tau$=$\tau$/ δt, where δt is the size of the time step used in the MD simulations. The calculated MSD (u$^2$) of individual oxygen ions in δ-Bi$_2$O$_3$ and in (Bi$_{0.7}$Y$_{0.3}$)$_2$O$_3$ at various temperatures is plotted in **Fig 5**. The calculated value of the MSD for δ-Bi$_2$O$_3$ at 1100 K at 20 ps is obtained to be about 13 Å$^2$ which is in agreement with reported results[1]. It has been found that at 1100 K, the value of MSD for δ-Bi$_2$O$_3$ is almost 5 times that of (Bi$_{0.7}$Y$_{0.3}$)$_2$O$_3$.

The displacement pattern of oxygen atoms in δ-Bi$_2$O$_3$ and (Bi$_{0.7}$Y$_{0.3}$)$_2$O$_3$ is shown in **Fig. 6**. We find that at 1000 K in δ-Bi$_2$O$_3$, oxygen ions have a distribution of jump lengths from 2 to 5 Å, while in (Bi$_{0.7}$Y$_{0.3}$)$_2$O$_3$ the maximum jump length is smaller and is about 4 Å. In (Bi$_{0.7}$Y$_{0.3}$)$_2$O$_3$ at a lower temperature of 700 K, oxygen ions have shorter jumps distances, as expected, ranging from 2 to 3 Å. The distribution of jump lengths is consistent with the distribution of first-neighbor oxygen-oxygen distances (**Fig. 4**) between 2.5 to 3.5 Å and the second neighbor distance at about 5 Å. The pattern of the MSD shows that oxygen atoms diffuse by jumping from one site to another. There is no specific pattern of displacement or any preference of direction as seen from the displacement pattern of a few selected oxygen ions in the disordered and in the doped phase. Further increase in temperature to 1100 K, increases the average MSD of oxygen's in δ-$Bi_2O_3$ and in $(Bi_{0.7}Y_{0.3})_2O_3$ by about 80% and 55%, respectively. This MSD in δ-$Bi_2O_3$ is more distributed than in $(Bi_{0.7}Y_{0.3})_2O_3$, predicting a range of possible jump distances in undoped δ-$Bi_2O_3$. This result is consistent with the sudden increase in diffusion coefficient of oxygen in δ-$Bi_2O_3$ beyond 1000 K (**Fig 7(a)**).

From the linear fit of the slope of the mean square displacements (**Fig 5**) of oxygen ions at different temperatures, diffusion coefficient of the two compounds have been computed as given in **Fig. 7 (a)**. The calculated self-diffusion coefficient of oxygen in cubic $Bi_2O_3$ is of the same order as reported by Wind et al[1]. The doping of yttrium in the lattice inhibits the free movement of oxygen atoms, which can be found with reduced diffusion coefficients. The diffusion coefficient of oxygen in $(Bi_{0.7}Y_{0.3})_2O_3$ at 1100 K is almost five times smaller ($2 \times 10^{-10}$ m$^2$/s) than in δ-$Bi_2O_3$, which has liquid like conductivity (~$10^{-9}$ m$^2$/s) reported previously[7]

In order to estimate the activation energy barriers in both compounds, the temperature dependence of diffusion coefficients is fitted with Arrhenius relation, i.e.,

$$D(T) = D_0 \exp(-E_a/k_B T) \qquad (4)$$

One can linearize this equation, i.e.,

$$\ln(D(T)) = \ln(D_0) - E_a/k_B T \qquad (5)$$

where $D_0$ is a constant factor representing diffusion coefficient at infinite temperature, $k_B$ is the Boltzmann constant and T is temperature in K. The calculated diffusion coefficients as a function of temperature is fitted (**Fig. 7(b)**) to equation (5). It can be found (see **Fig. 7(b)**) that the slope of δ-$Bi_2O_3$ is more negative than that of the doped $Bi_2O_3$, hence activation energy for oxygen diffusion is lower in yttria-doped $Bi_2O_3$ as compared to disordered δ-$Bi_2O_3$. Fitting of the data in **Fig. 7(b)** for δ-$Bi_2O_3$ gives a value of $9.4 \times 10^{-10}$

m$^2$/s for D$_0$ and the activation energy for oxygen diffusion is obtained as 0.87 eV, which is in excellent agreement with the value reported by Wind et al[1]. In the case of (Bi$_{0.7}$Y$_{0.3}$)$_2$O$_3$ we find (see **Fig 7(b)**) that there are two different sets of activation energies at two different ranges of the temperature. At temperature range 300 K - -700 K, the values of D$_o$ and activation energy are 8.1 × 10$^{-11}$ m$^2$/s and 0.04 eV respectively, while on increasing temperature beyond 700 K, these values are 7.5 × 10$^{-8}$ m$^2$/s and 0.58 eV respectively. This gives the inference that although the absolute value of diffusion in the Y doped Bi$_2$O$_3$ is lower than that of the pure cubic disordered phase, the activation energy is much lower near room temperature. The activation energy in (Bi$_{0.7}$Y$_{0.3}$)$_2$O$_3$ is almost 20 times lower than that in undoped δ- Bi$_2$O$_3$, hence it is easier for oxygen to start diffusing in the doped system as compared to the pure phase. Having said that, the presence of the larger size Y atoms poses a hindrance, hence lesser number of oxygen atoms diffuses in the doped phase in comparison to the disordered pure δ-phase. With increase in temperature, the activation energies become comparable, but lesser availability of free space compounded by the larger size of Y atoms do not allow diffusion of oxygen to reach as high values as they reach in the pure cubic Bi$_2$O$_3$.

**Figure 8 and 9** give the pathways in which the oxygen ions diffuse in both the compounds. Diffusion of selected oxygen ions over several picoseconds have been plotted; it can be seen that Bi forms a rigid framework while oxygen shows extensive diffusion. In case of the cubic δ-phase, the two selected oxygen ions appear to diffuse along b-direction and then along the c-direction. In case of the doped oxide, both selected oxygens diffuse within a plane and moving along an edge. There appears to be no preferred direction for diffusion, both in the pure and doped Bi$_2$O$_3$. These observations are in agreement with those of Wind et al[1], namely that jumps are essentially isotropic, not dominated by any preferred direction

From the QENS experiment, broadening of the elastic line is obtained from the diffusion of the oxygen in Bi$_2$O$_3$. As oxygen is a coherent scatterer, the microscopic understanding of the diffusive behavior of oxygen atoms in the lattice may be obtained by calculating the dynamical coherent structure factor for δ-Bi$_2$O$_3$. The lowest value of Q obtained from calculation is 0.55 Å$^{-1}$, which is limited by the size of the supercell.

The Lorentz peak functions were fitted to the calculated dynamical incoherent structure factors $S(Q, E)$

$$S(Q,E) = A1 \frac{\Gamma_1}{\Gamma_1^2 + E^2}$$

The Bi and O both scatter coherently, so the Q dependence of coherent $S(Q,E)$ has also been obtained from ab-initio calculation. The coherent $S(Q,E)$ was better described by combination of Lorentz and Gaussian peak functions, which are fitted to the calculated dynamical coherent structure factors $S(Q, E)$

$$S(Q,E) = \frac{2A1\,\Gamma_1}{4\pi[E^2+\Gamma_1^2]} + \frac{A2}{\Gamma_2\sqrt{\pi/2}} exp^{-(\frac{2E^2}{\Gamma_2^2})} \qquad (6)$$

Here A1 and A2 are the area of Lorentzian and Gaussian respectively, while $\Gamma_1$ and $\Gamma_2$ are the half width at half maximum energy (HWHM) of the Lorentzian and the standard deviation of the Gaussian, respectively. The sum of the area of Lorentzian and Gaussian gives the static structure factor, $S(Q)$. In the fitting procedure the value of $\Gamma_2$ is fixed at 0.027 meV, which corresponds to the resolution of the OSIRIS spectrometer. The same value of the resolution was used while calculating the coherent $S(Q, E)$ from AIMD calculations. The extracted Q dependence of HWHM from coherent and incoherent calculations at 1100 K is shown in **Fig. 10(b).** The areas of the Lorentzian and Gaussian from coherent calculations at 1100 K are also shown in **Fig. 10(c)**.

The Q-dependence of the HWHM shows an oscillatory behavior due to the combined effect of the jump diffusion and coherent scattering. The analysis of the QENS data of $\delta$-$Bi_2O_3$ reported by Wind et al indicated[1] that coherency effects, i.e., the structure factor, must be included to analyse the experimental data of HWHM vs Q, using the so-called Skold modification[48] to the Chudley-Elliott model[52]

$$\Gamma(Q) \times S(Q) = (1 - Sin(Qd)/Qd) / \tau \qquad (7)$$

where d is the jump distance and τ is the relaxation time. We have also plotted (**Fig 10(d)**) the multiplication of S(Q) and HWHM of the Lorentzian. We find it difficult to fit the above equation (7) to the calculations (**Fig 10(d)**). Although our calculations are in qualitative agreement with our experimental observations as well as that of Wind et al[1], our analysis suggests that diffusion in δ-$Bi_2O_3$ cannot be described by a single jump diffusion model.

As shown in **Fig. 10(b),** we find that the nature of dependence of HWHM on Q obtained from incoherent S(Q,E) matches well with the experimental data (**Fig. 10(a)**), although the values of the calculated HWHM are lower in comparison to those of the experimental data. This result predicts that in the presence of 25% oxygen vacancies in the structure of δ-phase of $Bi_2O_3$, oxygen may move rather independently. Therefore, the oxygen dynamics in this material may be described by the incoherent approximation.

A combination of two incoherent jump diffusion lengths ranging from 3.3 Å – 10.8 Å and relaxation time 4.1-5.1 ps in the form of the pure incoherent Chudley-Elliott model[52] can be fitted well with experimental HWHM (**Fig. 11**). In this fitting two different Q ranges have been fitted with two different set of parameters. For low Q, .i.e., below 1.3 Å$^{-1}$ about 80% of all O ions have the jump distance of 10.8 Å and relaxation time of 5.1 ps, however, for high Q , i.e., above 1.3 Å$^{-1}$ , jump distance and relaxation time of all O ions are 3.3 Å and 4.1 ps, respectively. From Eq. (7) it is clear that for large Q, when S(Q) can be normalized to unity, the Skold modification[48] reduces to the incoherent Chudley-Elliott model[52]. This result predicts again that the oxygen vacancies in the δ-$Bi_2O_3$ play a crucial role in its diffusion process. In the absences of 25% oxygen in δ-$Bi_2O_3$, as explained above, oxygen can diffuse more independently and thus motion can be well described with incoherent approximations. This finding agrees well with earlier reports of incoherent diffusion in δ-$Bi_2O_3$ [15]. The average diffusion coefficient is thus obtained as 2.14 × $10^{-8}$ m$^2$/s which compares well with calculated results given above and reported values.[1]

## V. CONCLUSIONS

In this article we have reported inelastic neutron scattering data and computational simulations on $Bi_2O_3$ in the ambient and high temperature phases, and on Y-stabilized $Bi_2O_3$ ($Bi_{0.7}Y_{0.3}$)$_2O_3$. Our ab-initio lattice dynamics results based on DFT are in very good agreement with the measured data at ambient conditions.

High temperature VDOS is measured to understand the increased disorder in the lattice. With increasing temperature, beyond 1000 K, a complete disappearance of sharp vibrational peaks in the measured data indicates the appearance of the highly disordered cubic δ-phase of $Bi_2O_3$. Our AIMD calculations of the VDOS indicate that O contributes in the entire frequency spectrum while Bi and Y (in doped $Bi_2O_3$) contribute only in the lower energy transfer range. Pair correlations between Bi-O, Bi-Bi, O-O suggest that only the oxygen sub-lattice becomes disordered in the high temperature δ-phase. Calculated MSD of oxygen in δ-phase of $Bi_2O_3$ show that oxygen moves randomly and isotropically from one site to the another and there is no specific preference of direction. Our calculations assume random ordering in the δ-phase which is partially vacant. A combination of different sets of jump distances and relaxation times, obtained from QENS experiments, are determined for different ranges of momentum transfer vectors. Two broad range of parameters are observed for diffusion, for low Q, i.e., below 1.3 $Å^{-1}$ the jump distance and relaxation time are of 10.8 Å and 5.1 ps, respectively. However, for high Q, i.e., above 1.3 $Å^{-1}$, jump distance and relaxation time are of 3.3 Å and 4.1 ps, respectively.

The pair correlations in δ-phase and in $(Bi_{0.7}Y_{0.3})_2O_3$ are similar emphasizing the fact that high conducting cubic phase can be stabilized at lower temperature by Y doping. Our calculations indicate that diffusion coefficient of oxygen atoms decrease by one-fifth in $(Bi_{0.7}Y_{0.3})_2O_3$ in comparison to the liquid like conduction of the cubic δ-phase, yet increases the conductivity considerably compared to the ambient pure α-phase $Bi_2O_3$ which does not show any conduction.

## ACKNOWLEDGEMENTS

We thank the ISIS Neutron and Muon Source Facility for the provision of beam time (RB1410490). Institut Laue-Langevin (ILL) facility, Grenoble, France, is acknowledged for providing beam time on the IN4C spectrometer. The use of ANUPAM super-computing facility at BARC is acknowledged. SLC thanks the Indian National Science Academy for award of an INSA Senior Scientist position.

[1] J. Wind, R. A. Mole, D. Yu, and C. D. Ling 2017 *Chemistry of Materials* **29**, 7408
[2] T. Takahashi and H. Iwahara 1978 *Materials Research Bulletin* **13**, 1447
[3] C. E. Mohn, S. Stølen, S. T. Norberg, and S. Hull 2009 *Physical Review B* **80**, 024205
[4] C. E. Mohn, S. Stølen, S. T. Norberg, and S. Hull 2009 *Physical Review Letters* **102**, 155502
[5] N. Sammes, G. Tompsett, H. Näfe, and F. Aldinger 1999 *Journal of the European Ceramic Society* **19**, 1801
[6] A. Seko, Y. Koyama, A. Matsumoto, and I. Tanaka 2012 *Journal of Physics: Condensed Matter* **24**, 475402
[7] R. D. Bayliss, S. N. Cook, S. Kotsantonis, R. J. Chater, and J. A. Kilner 2014 *Advanced Energy Materials* **4**, 1301575
[8] R. Packer and S. Skinner 2010 *Advanced Materials* **22**, 1613
[9] M. Burriel, G. Garcia, J. Santiso, J. A. Kilner, R. J. Chater, and S. J. Skinner 2008 *Journal of materials chemistry* **18**, 416
[10] R. Packer, S. Skinner, A. Yaremchenko, E. Tsipis, V. Kharton, M. Patrakeev, and Y. A. Bakhteeva 2006 *Journal of materials chemistry* **16**, 3503
[11] S. J. Skinner and J. A. Kilner 2000 *Solid State Ionics* **135**, 709
[12] M. Verkerk and A. Burggraaf 1981 *Journal of The Electrochemical Society* **128**, 75
[13] H. Harwig and A. Gerards 1978 *Journal of Solid State Chemistry* **26**, 265
[14] D. S. Aidhy, J. C. Nino, S. B. Sinnott, E. D. Wachsman, and S. R. Phillpot 2008 *Journal of the American Ceramic Society* **91**, 2349
[15] S. Hull, S. T. Norberg, M. G. Tucker, S. G. Eriksson, C. E. Mohn, and S. Stølen 2009 *Dalton Transactions*, 8737
[16] G. Gattow and H. Schröder 1962 *Zeitschrift für anorganische und allgemeine Chemie* **318**, 176
[17] X. Gong, J. Huang, Y. Chen, M. Wu, G. Liu, X. Lei, J. Liang, H. Cao, F. Tang, and B. Xu 2013 *Int. J. Electrochem. Sci* **8**, 10549
[18] P. Shuk, H. D. Wiemhöfer, U. Guth, W. Göpel, and M. Greenblatt 1996 *Solid State Ionics* **89**, 179
[19] E. Mamontov 2016 *Solid State Ionics* **296**, 158
[20] P. Battle, C. Catlow, and L. Moroney 1987 *Journal of Solid State Chemistry* **67**, 42
[21] G. Malmros 1970 *Acta Chemica Scandinavica* **24**, 384Ā396
[22] S. Boyapati, E. D. Wachsman, and B. C. Chakoumakos 2001 *Solid State Ionics* **138**, 293
[23] S. Sanna, V. Esposito, J. W. Andreasen, J. Hjelm, W. Zhang, T. Kasama, S. B. Simonsen, M. Christensen, S. Linderoth, and N. Pryds 2015 *Nature materials* **14**, 500
[24] F. Schröder, N. Bagdassarov, F. Ritter, and L. Bayarjargal 2010 *Phase Transitions* **83**, 311
[25] T. Willis, D. Porter, D. Voneshen, S. Uthayakumar, F. Demmel, M. Gutmann, M. Roger, K. Refson, and J. Goff 2018 *Scientific reports* **8**, 1
[26] K. Funke 1991 *Philosophical Magazine A* **64**, 1025
[27] G. Lucazeau, J. Gavarri, and A. Dianoux 1987 *Journal of Physics and Chemistry of Solids* **48**, 57
[28] M. Gupta, B. Singh, P. Goel, R. Mittal, S. Rols, and S. Chaplot 2019 *Physical Review B* **99**, 224304
[29] B. Singh, M. K. Gupta, R. Mittal, M. Zbiri, S. Rols, S. J. Patwe, S. N. Achary, H. Schober, A. K. Tyagi, and S. L. Chaplot 2017 *Physical Chemistry Chemical Physics* **19**, 15512
[30] B. Singh, M. K. Gupta, R. Mittal, and S. L. Chaplot 2018 *Journal of Materials Chemistry A* **6**, 5052
[31] B. Singh, M. K. Gupta, R. Mittal, M. Zbiri, S. Rols, S. J. Patwe, S. N. Achary, H. Schober, A. K. Tyagi, and S. L. Chaplot 2017 *Journal of Applied Physics* **121**, 085106
[32] S. Mukhopadhyay and F. Demmel, in *AIP Conference Proceedings* (AIP Publishing LLC, 2018), p. 030001.
[33] F. Demmel and S. Mukhopadhyay 2016 *The Journal of Chemical Physics* **144**, 014503

[34] K. S. D. L. Price *Neutron scattering* (Academic Press, Orlando, 1986), Vol. A.
[35] J. M. Carpenter and D. L. Price 1985 *Physical Review Letters* **54**, 441
[36] S. Rols, H. Jobic, and H. Schober 2007 *Comptes Rendus Physique* **8**, 777
[37] F. Demmel, D. McPhail, J. Crawford, D. Maxwell, K. Pokhilchuk, V. Garcia-Sakai, S. Mukhopadyay, M. Telling, F. Bermejo, and N. Skipper, in *EPJ Web of Conferences* (EDP Sciences, 2015), p. 03003.
[38] S. Mukhopadhyay, B. Hewer, S. Howells, and A. Markvardsen 2019 *Physica B: Condensed Matter* **563**, 41
[39] H. Harwig 1978 *Zeitschrift für anorganische und allgemeine Chemie* **444**, 151
[40] G. Kresse and J. Furthmüller 1996 *Computational Materials Science* **6**, 15
[41] G. Kresse and D. Joubert 1999 *Physical Review B* **59**, 1758
[42] J. P. Perdew, K. Burke, and M. Ernzerhof 1997 *Physical Review Letters* **78**, 1396
[43] J. P. Perdew, K. Burke, and M. Ernzerhof 1996 *Physical Review Letters* **77**, 3865
[44] H. J. Monkhorst and J. D. Pack 1976 *Physical Review B* **13**, 5188
[45] P. Software and K. Parlinksi, PHONON Software, 2003.
[46] S. Nosé 1984 *The Journal of chemical physics* **81**, 511
[47] G. R. Kneller, V. Keiner, M. Kneller, and M. Schiller 1995 *Computer physics communications* **91**, 191
[48] K. Sköld 1967 *Physical Review Letters* **19**, 1023
[49] S. F. Parker, S. Mukhopadhyay, M. Jiménez-Ruiz, and P. W. Albers 2019 *Chemistry – A European Journal* **25**, 6496
[50] M. P. Allen and D. J. Tildesley, *Computer simulation of liquids* (Oxford university press, 2017).
[51] A. K. Sagotra, D. Chu, and C. Cazorla 2019 *Physical Review Materials* **3**, 035405
[52] C. T. Chudley and R. J. Elliott 1961 *Proceedings of the Physical Society* **77**, 353

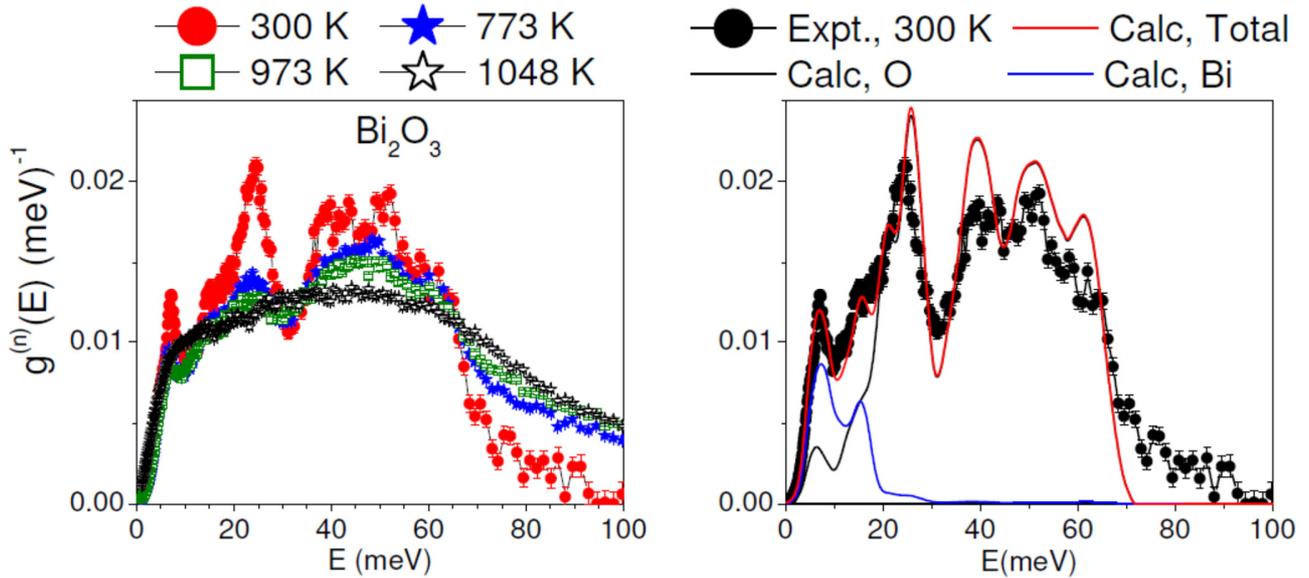

FIG 1(Color online) Inelastic neutron scattering spectra (line with symbols) measured $Bi_2O_3$ at 300 K, 773 K, 973 K and at 1048 K compared with our ab-initio lattice dynamical calculation in the ordered monoclinic α-phase at 0 K. The computed total neutron-weighted phonon density of states as well as the partial components due to the oxygen and bismuth atoms are shown in the right panel.

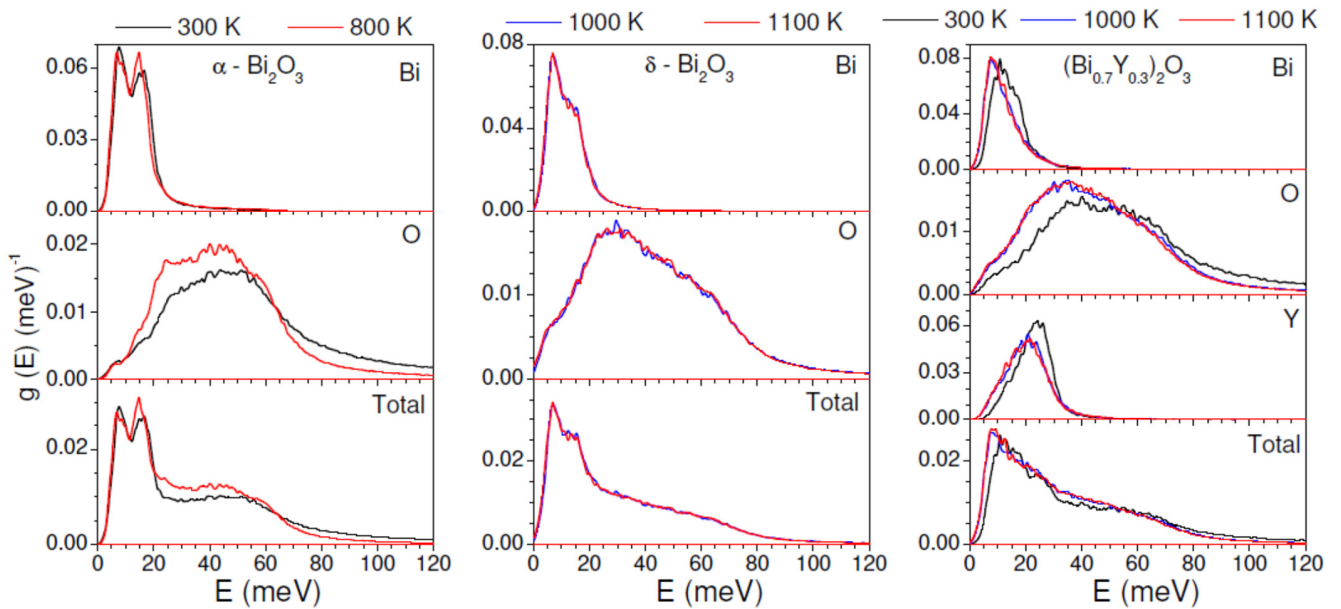

FIG 2 (Color online) Computed partial and total phonon density of states of different phases of $Bi_2O_3$ and Y-doped $Bi_2O_3$ at different temperature from ab-initio MD calculations.

FIG 3 (Color online) Inelastic neutron scattering spectra (line with symbols) measured at 300 K and at higher temperature, 1048 K compared with our computed phonon density of states using ab-initio MD at 300 K and 1100 K.

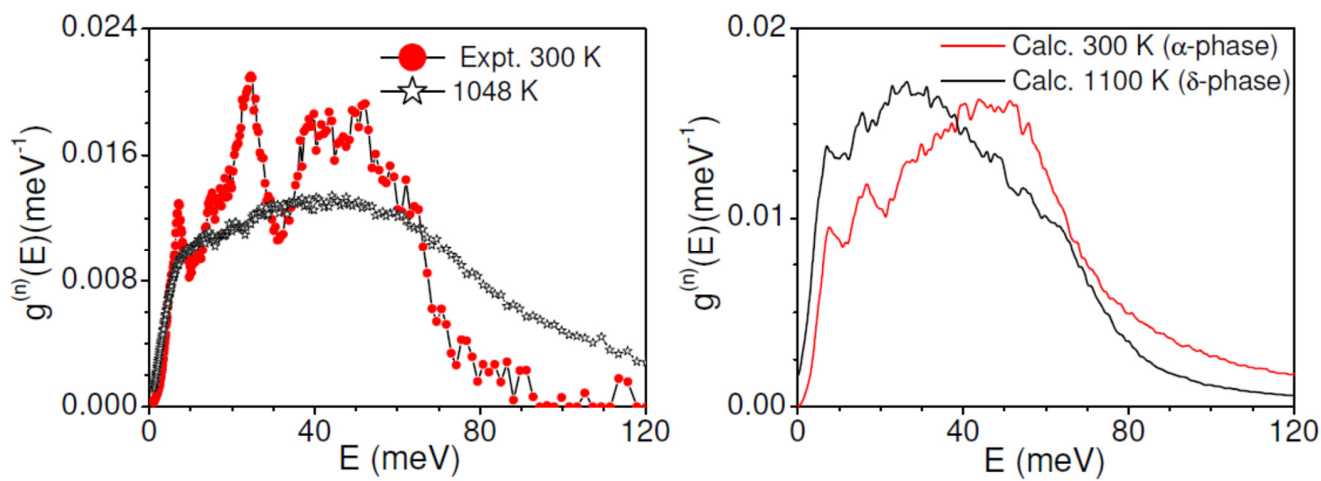

FIG 4 (Color online) Computed pair correlations of different atom pairs and total G(r,r') in α-$Bi_2O_3$, δ-$Bi_2O_3$ and Y-doped $Bi_2O_3$ at different temperatures from ab-initio MD.

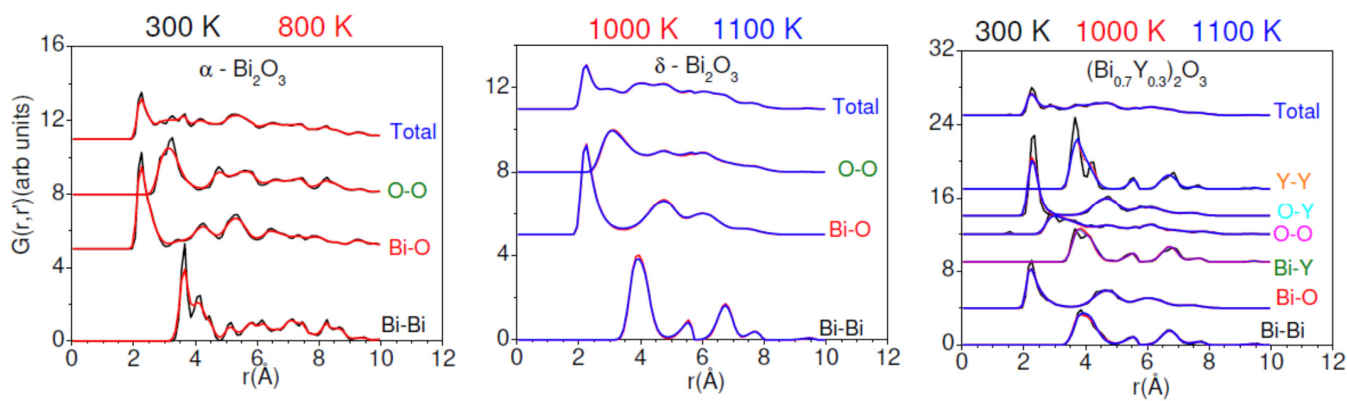

FIG. 5 (Color online) Mean square displacement ($<u^2>$) of oxygen ions in $\delta$-$Bi_2O_3$ and $(Bi_{0.7}Y_{0.3})_2O_3$. Green line is a linear fit to the computed data at different temperatures.

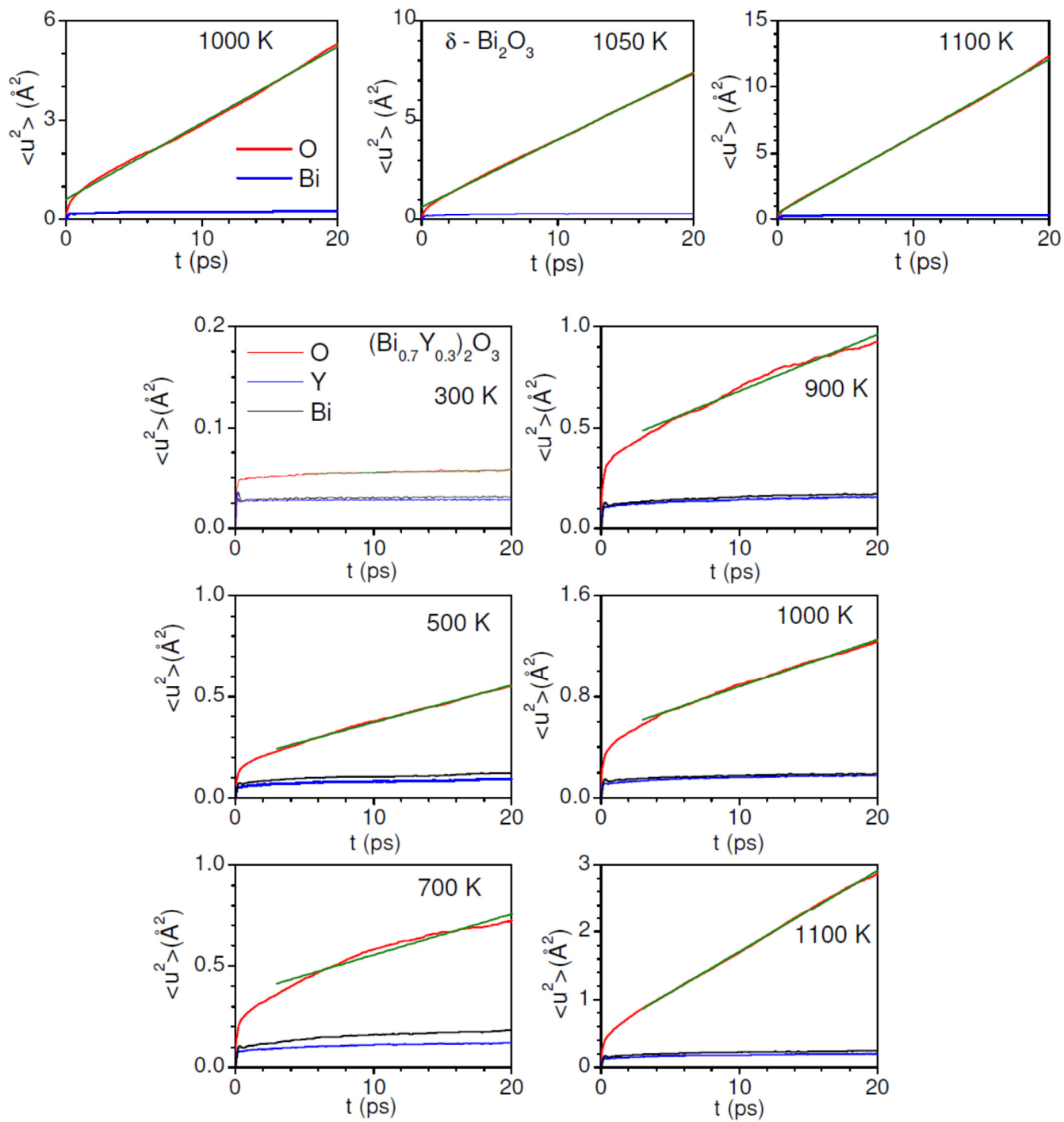

FIG 6 (Color online) The calculated mean squared displacements ($<u^2>$) of individual oxygen atoms in the $\delta - Bi_2O_3$ and $(Bi_{0.7}Y_{0.3})_2O_3$. The $<u^2>$ of selected oxygen atoms at 1000 K is shown with thick solid lines. The trajectories of these oxygen atoms are shown in Figs 8 and 9.

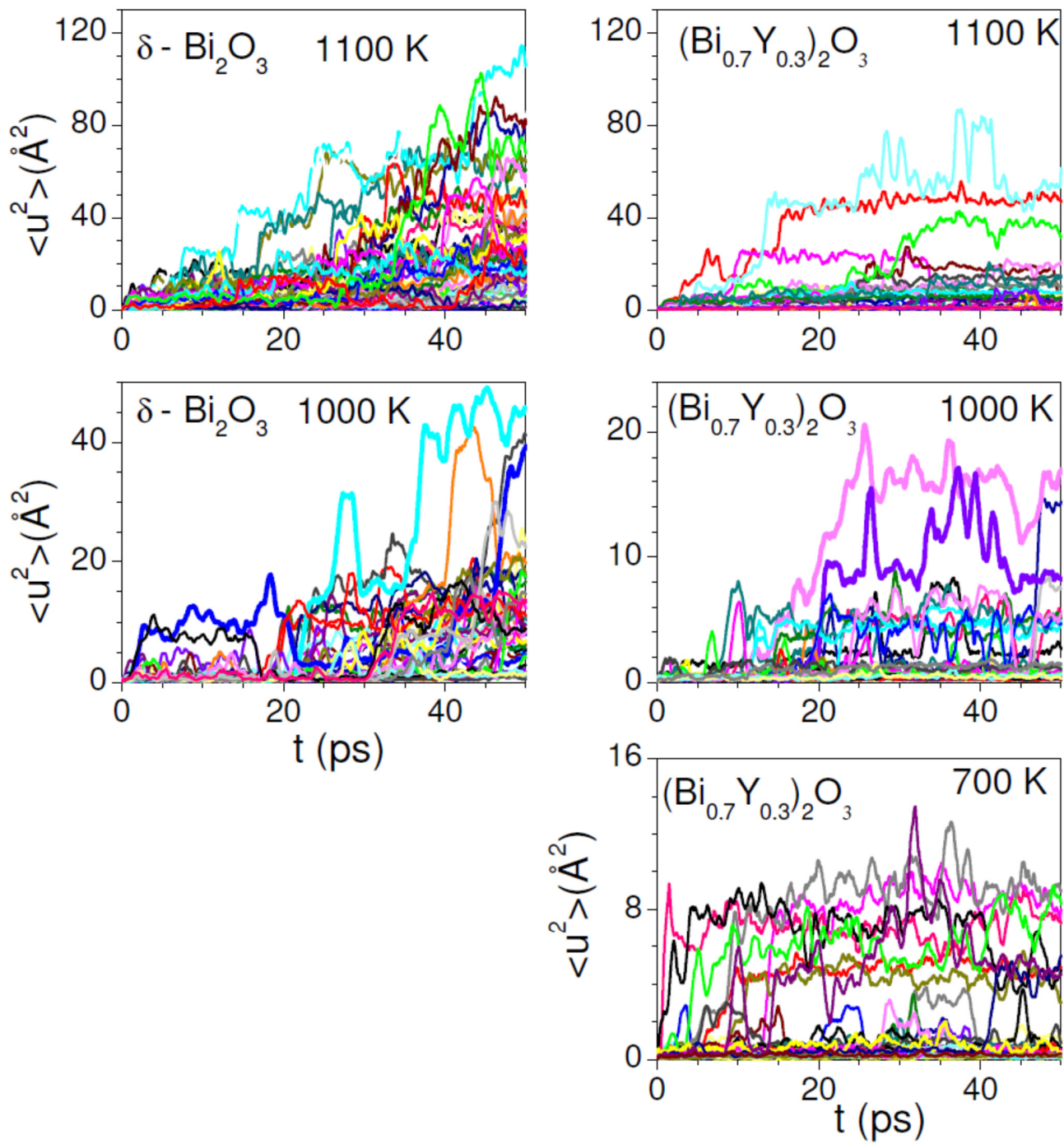

FIG 7 (Color online) Diffusion coefficient and activation energy barriers of oxygen in δ - $Bi_2O_3$ and $(Bi_{0.7}Y_{0.3})_2O_3$ from ab-initio MD simulations.

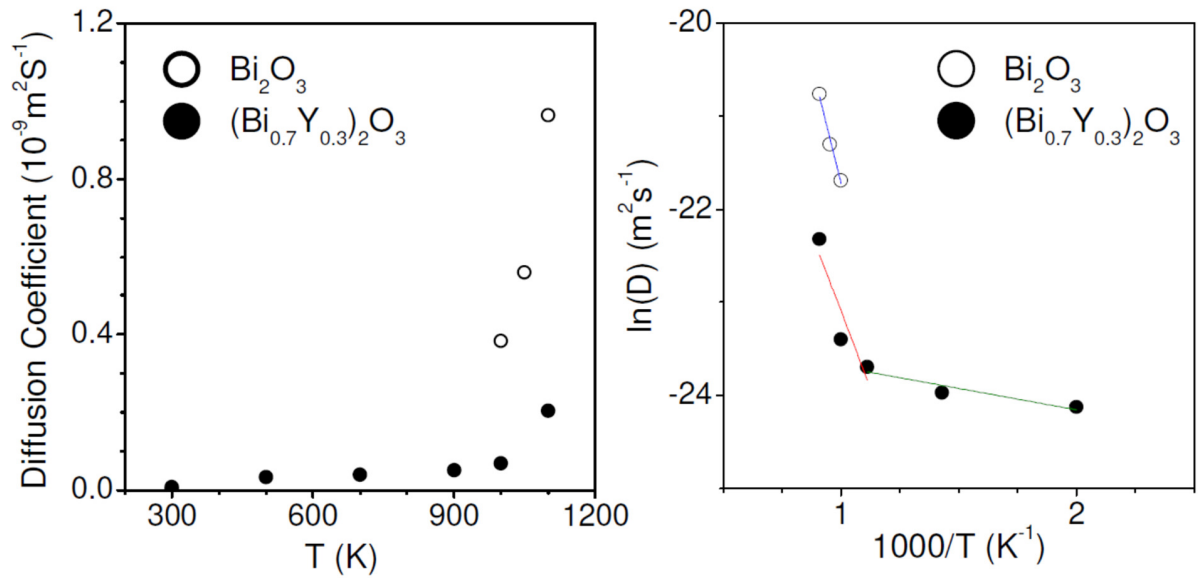

FIG 8 (Color online) Calculated trajectories of selected O atoms at 1000 K in δ - $Bi_2O_3$. Trajectories of two different oxygen atoms are shown. Red spheres represent oxygen atoms at their lattice sites, blue ones are the Bi atoms. The time dependent positions of the selected oxygen atoms are shown by green colored dots.

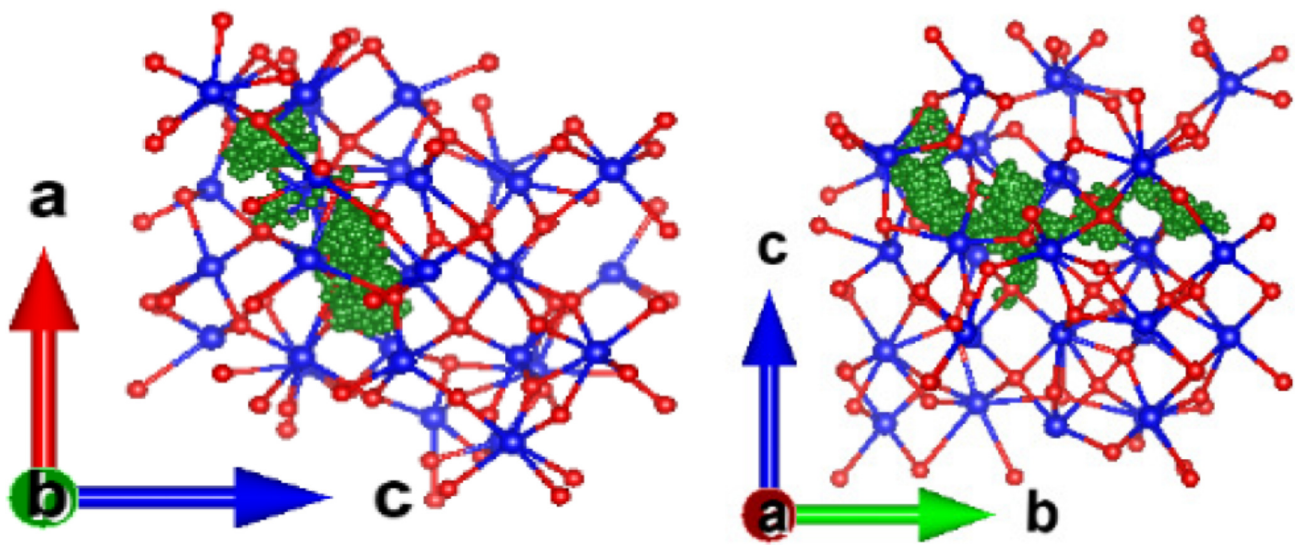

Fig. 9: Calculated trajectories of selected oxygen ions in $(Bi_{0.7}Y_{0.3})_2O_3$ at 1000 K. Red spheres represent oxygen atoms at their lattice sites, blue and yellow ones are the Bi and Y atoms respectively. The time dependent positions of the selected oxygen atoms are shown by green colored dots.

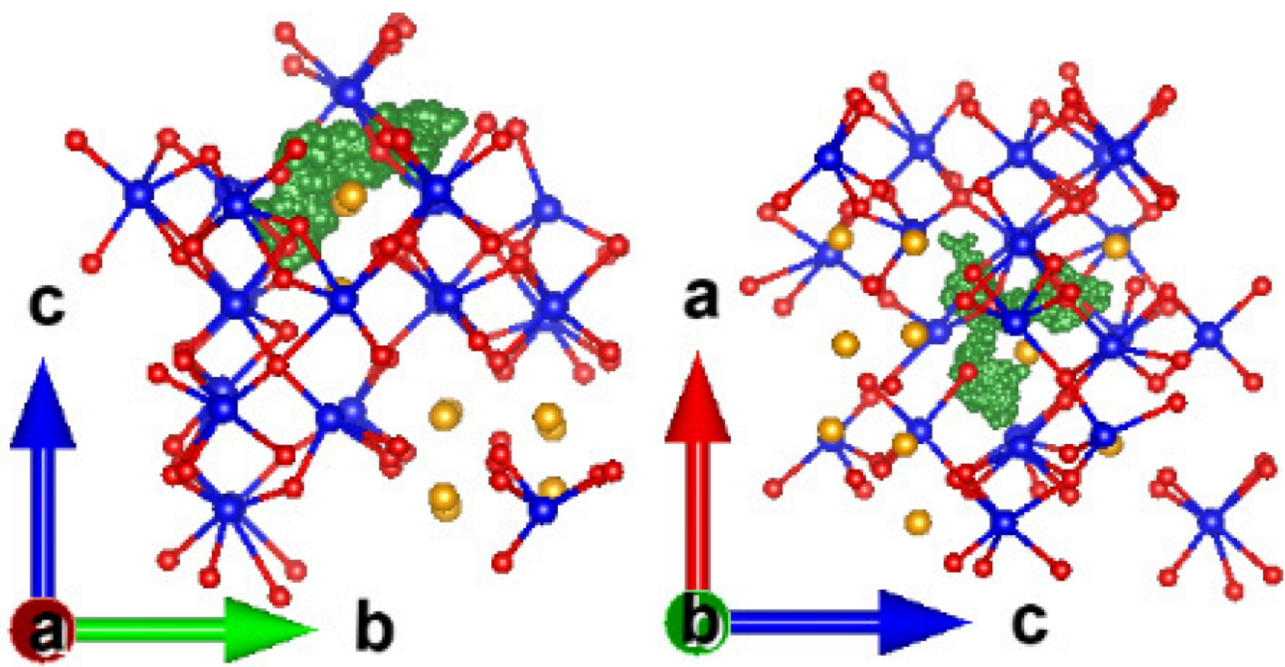

FIG 10. (a) The variation of wave vector dependence (Q) of half width at half maximum (HWHM) of Lorentzian peak fitted to the experimental dynamical neutron scattering function S(Q,E) obtained from QENS experiment for $\delta$ – $Bi_2O_3$ at 1083 K. (b) The calculated Q dependence of HWHM of Lorentzian obtained from fitting to coherent (open circles) and incoherent (closed circles) S(Q,E) obtained from ab-initio calculation. (c) The area of Gaussian and Lorentzian extracted from fitting to coherent S(Q,E) from ab-initio calculation. The full width at half maximum (FWHM) of Gaussian has been fixed at 0.027 meV, which corresponds to the resolution of the OSIRIS spectrometer. S(Q) is normalized to 1 at 1.6 Å$^{-1}$ which is below the first Bragg peak. (d) The multiplication of the HWHM of Lorentzian and sum of the area of Gaussian and Lorentzian.

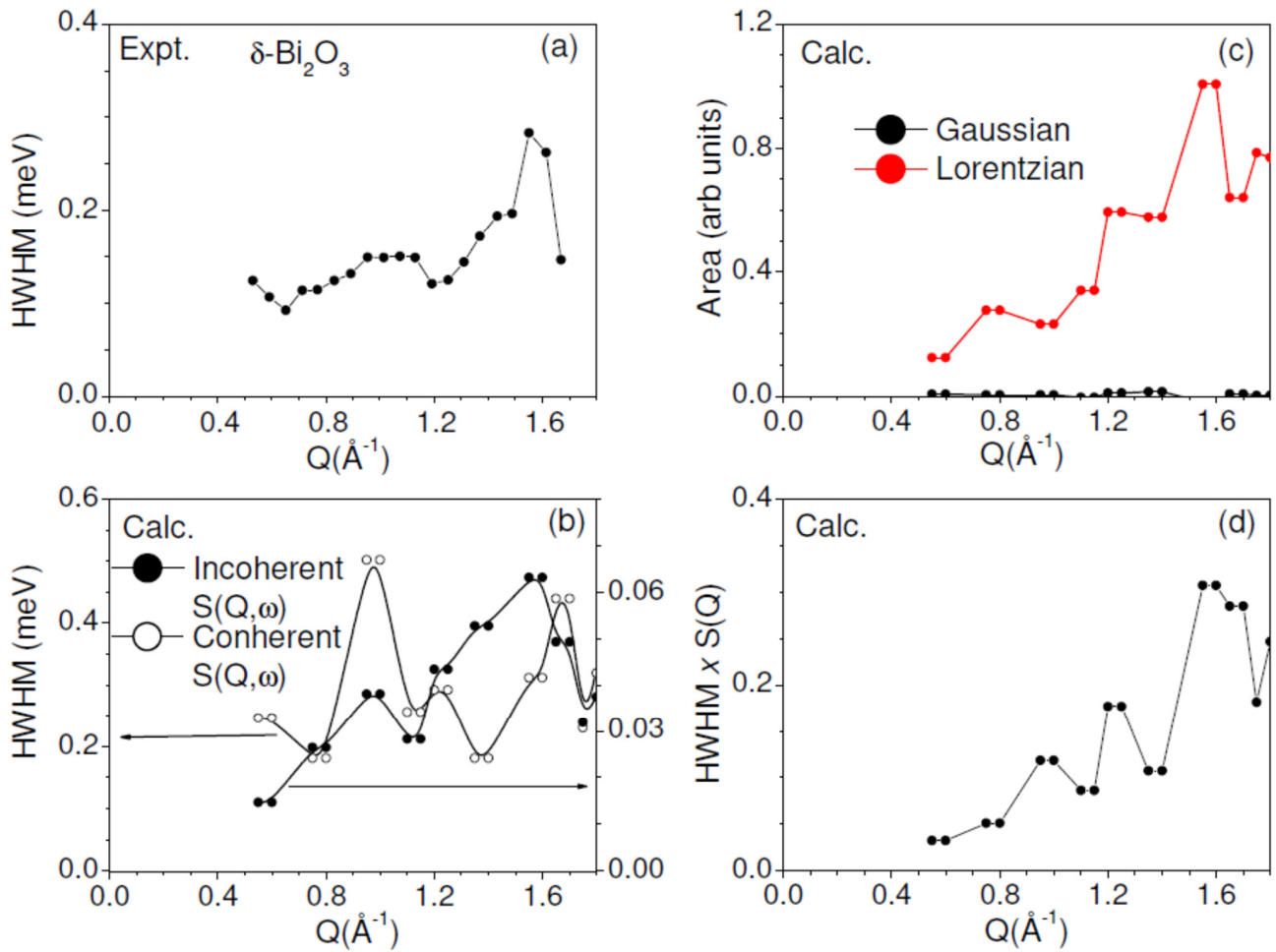

FIG 11. The analysis of QENS results using incoherent Chudley-Elliott model. (a) Normalized QENS spectra for Q = 1.1 A, (b) fitting of HWHM (Q) of QENS spectra at 1083 K using Chudley-Elliott model[52] with two sets of parameters. See text for more details.

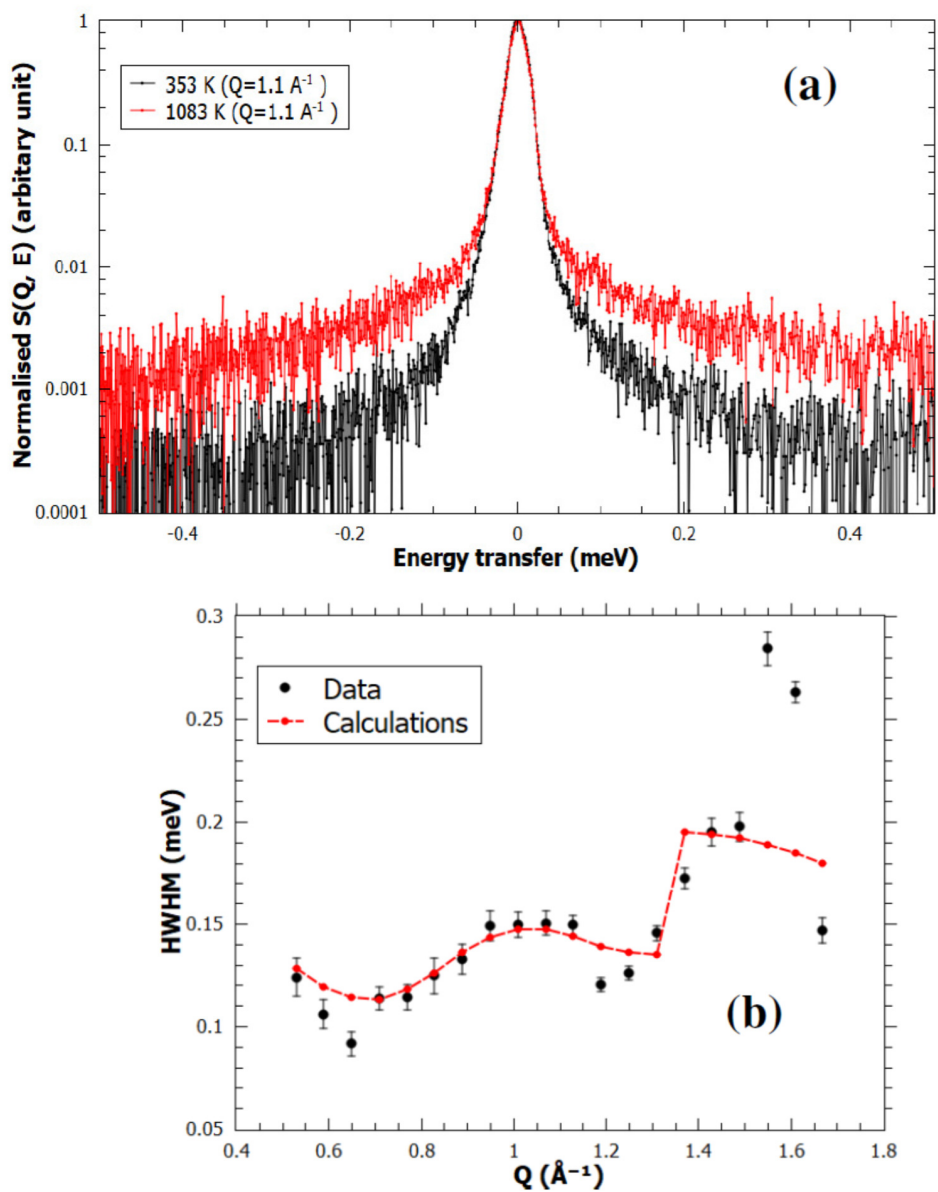